%% file: main.tex
\documentclass[letterpaper,twocolumn,10pt]{article}
\usepackage{usenix2019_v3}
\input{macros}

\usepackage{tikz}

\usepackage[normalem]{ulem}
\usepackage{microtype}
\usepackage[linesnumbered,algoruled,boxed,lined]{algorithm2e}
\usepackage{scalerel}
\usepackage{amsmath}
\usepackage{amsfonts}
\usepackage[euler]{textgreek}

\usepackage{algcompatible}
\usepackage{algpseudocode}
\usepackage{setspace}
\usepackage{booktabs}
\usepackage{array}
\usepackage{enumitem}
\usepackage{upgreek}
\usepackage{multirow}
\usepackage{multicol}
\usepackage{float}
\usepackage{adjustbox}
\usepackage{tabularx}
\usepackage{longtable}
\usepackage{changepage}
\usepackage{relsize}
\usepackage{courier}
\usepackage{ifthen}
\usepackage[sc]{caption}

\begin{document}

\ifdefined\ifrevision
	\pagenumbering{Alph}
	\input{revision-log}
	\pagenumbering{arabic}
\fi

\ifdefined\ifsubmission
\pagenumbering{gobble}
\fi

\title{\Large \bf \sys: Hybrid Side-Channel-Resilient Caches\\ for Trusted Execution Environments} 

\author{
\rm{Ghada Dessouky,
Tommaso Frassetto,
Ahmad-Reza Sadeghi}\\
Technische Universit\"at Darmstadt, Germany\\
\tt{\{ghada.dessouky, tommaso.frassetto, ahmad.sadeghi}\}\tt{@trust.tu-darmstadt.de}}
%%%%%%%%%%%%%%%%%%%%%%%%%%%%%%%%%%%%

\maketitle

\begin{abstract}\label{sec:abstract}
Modern multi-core processors share cache resources for maximum cache utilization and performance gains. However, this leaves the cache vulnerable to side-channel attacks, where inherent timing differences in shared cache behavior are exploited to infer information on the victim's execution patterns, ultimately leaking private information such as a secret key.
The root cause for these attacks is mutually distrusting processes sharing the cache entries and accessing them in a deterministic and consistent manner.  Various defenses against cache side-channel attacks have been proposed. However, they suffer from serious shortcomings: they either degrade performance significantly, impose impractical restrictions, or can only defeat certain classes of these attacks.
More importantly, they assume that side-channel-resilient caches are required for the entire execution workload and do not allow the possibility to selectively enable the mitigation only for the security-critical portion of the workload.

We present a generic mechanism for a flexible and soft partitioning of set-associative caches and propose a hybrid cache architecture, called \sys. \sys can be configured to selectively apply side-channel-resilient cache behavior only for isolated execution domains, while providing the non-isolated execution with conventional cache behavior, capacity and performance.  An isolation domain can include one or more processes, specific portions of code, or a Trusted Execution Environment (e.g., SGX or TrustZone). We show that, with minimal hardware modifications and kernel support, \sys can provide side-channel-resilient cache only for isolated execution with a performance overhead of 3.5--5\%, while incurring no performance overhead for the remaining execution workload. We provide a simulator-based and hardware implementation of \sys to evaluate the performance and area overheads, and show how \sys mitigates typical access-based and contention-based cache attacks.

\end{abstract}
\input{introduction}

\input{background}  
\input{attacks}

\input{design} 
\input{security}

\input{evaluation}
\input{discussion}
\input{related}
\input{conclusion} 
\input{ack}

%%%%%%% -- PAPER CONTENT ENDS -- %%%%%%%%

%%%%%%%%% -- BIB STYLE AND FILE -- %%%%%%%%
{\small
\bibliographystyle{plain}
\bibliography{main_bib}}
%%%%%%%%%%%%%%%%%%%%%%%%%%%%%%%%%%%%
\null  %to prevent the indentation error with the last line of the last reference
\end{document}

%% file: macros.tex
\def\ifsubmission{1}

\ifdefined\ifsubmission
  \newcommand{\TODO}[1]{}
\else
  \newcommand{\TODO}[1]{\marginpar{\textcolor{red}{TODO: #1}}}
\fi

\ifdefined\ifrevision
	\newcommand{\CHANGED}[1]{\textcolor{blue}{#1}}
  \newenvironment{CHANGeD}{\color{blue}\ignorespaces}{\ignorespacesafterend}
\else
  \newcommand{\CHANGED}[1]{#1}
  
\fi

\ifdefined\

\usepackage{xspace}
\newcommand{\sys}{\textsc{HybCache}\xspace}
\newcommand{\sd}{I-Domain\xspace}
\newcommand{\sds}{I-Domains\xspace}
\newcommand{\nsd}{NI-Domain\xspace}

\newcommand{\nsecure}{\textit{n$_{isolated}$}\xspace}

\newcommand{\subcache}{\textit{subcache}\xspace}
\newcommand{\ididlong}{\ensuremath{\mathtt{Isolation Domain ID}}\xspace}
\newcommand{\idid}{\ensuremath{\mathtt{IDID}}\xspace}
\newcommand{\idids}{\ensuremath{\mathtt{IDIDs}}\xspace}

\newcommand*\circled[1]{\tikz[baseline=(char.base)]{
            \node[shape=circle,draw,inner sep=.6pt] (char) {#1};}}

%% file: introduction.tex
%!TEX root = main.tex

\section{Introduction} \label{sec:intro}
For decades now, upcoming processor generations are being augmented with novel performance-enhancing capabilities.
Performance and security of processor architectures and microarchitectures are considered exclusively independent design metrics, with architects primarily focused on the more tangible performance benefits. 
However, the recent outbreak of micro-architectural cross-layer attacks~\cite{Kocher18,google-zero, Lipp18, koruyeh2018spectre, Kiriansky18, maisuradze2018ret2spec, cachebleed2017yarom, gras2018translation, evtyushkin2018branchscope, evtyushkin2016jump,lee2017inferring,aciiccmez2007predicting, aciiccmez2007power, clkscrew2017tang, memjam2018moghimi, foreshadow}, has demonstrated the critical and long-ignored effects of micro-architectural performance optimizations on systems from a security standpoint.
It is becoming evident how performance and security are at conflict with each other unless architects address the design trade-off early on and not as an afterthought.

One prominent performance feature and the subject of a wide range of recent architectural attacks is the use of caches and cache-like structures to provide orders-of-magnitude faster memory accesses. %that are orders of magnitude faster than off-chip DRAM memory accesses. 
The intrinsic timing difference between a cache hit and miss is one of various \emph{side channels} that can be exploited by an adversary process via a carefully crafted side-channel attack to infer the memory access patterns of a victim process~\cite{Gullasch11, Yarom14, Gruss16, Irazoqui16, Lipp16, Irazoqui15, Kayaalp16, Liu15, Gras2017, Osvik2006, Gruss15, Guanciale16, Yan19, VanSchaik18}.
Consequently, the adversary can leak unauthorized information, such as a private key, hence violating the confidentiality and isolation of the victim process.

\paragraph{Cache Side-Channel Attacks.} In earlier years, cache side-channel attacks have been shown to compromise cryptographic implementations~\cite{bernstein2005cache, Liu15, Osvik2006, Yarom14}. More recently, attack variants such as \emph{Prime + Probe}~\cite{Osvik2006, Irazoqui15, Kayaalp16, Liu15} and \emph{Flush + Reload} attacks~\cite{Gullasch11,Yarom14} are being demonstrated on a much larger scale. They have been shown to bypass address space layout randomization (ASLR)~\cite{Lipp16, Gras2017}, infer keystroke behavior~\cite{Gruss16,Gruss15}, or leak privacy-sensitive human genome indexing computation~\cite{Brasser17}, whereby millions of platforms using various architectures have been shown vulnerable to such attacks.
The attacks require an adversary to orchestrate particular cache evictions of target memory addresses of interest and after a time interval measure its own memory access latencies or observe relevant computation and profile how it has been affected. This enables the adversary to deduce the victim's memory access patterns and infer dependent secrets. 
Cache side-channel attacks have been shown to exploit core-specific caches as well as shared last-level caches across different cores or virtual machines~\cite{Kayaalp16,Liu15,Gruss15}.
Even hardware-security extensions and trusted execution environments (TEEs) such as Intel SGX~\cite{intel-sgx1, intel-sgx2} and ARM TrustZone~\cite{trustzone} are not immune to these attacks.
While they do not claim cache side-channel security, recent cache side-channel attacks targeting SGX~\cite{Brasser17,Schwarz17, Moghimi17, GES17} and TrustZone~\cite{ARMageddon,Zhang16} have been shown to compromise the acclaimed privacy and isolation guarantees of these security architectures, thus undermining their very purpose.

\paragraph{Existing Cache Defenses.} To defeat cache side-channel attacks, there has been extensive research on techniques to identify and mitigate information leaks in a software's memory access patterns~\cite{Doychev15,Doychev17,Kopf12}. However, mitigating these leaks efficiently for arbitrary software (beyond cryptographic implementations) remains impractical and challenging.
Alternatively, hardware-based and software approaches have been proposed to modify the cache organization itself to limit cache interference across different security domains.
Examples include modifying replacement and leveraging inclusion policies~\cite{Kayaalp17, Yan17}, as well as approaches that rely on cache partitioning~\cite{Wang16, Zhou16, Kiriansky17, Kim12, Liu16, Gruss17, Wang07}, and randomization/obfuscation-based schemes~\cite{Wang07, Newcache16, Liu14,  Trilla18, Qureshi18} to randomize the relation between the memory address and its cache set index.

While strict cache partitioning is the intuitive approach to provide complete cache isolation and non-interference between mutually distrusting processes, it remains highly impractical and prevents efficient cache utilization. 
On the other hand, randomization-based approaches make the attacks computationally much more difficult by randomizing the mapping of memory addresses to cache sets.
However, existing schemes either require complex management logic, impose particular restrictions, rely on weak cryptographic functions, or mitigate only some classes of cache side-channel attacks.
Most importantly, all of the aforementioned schemes are designed to provide side-channel cache protection for the entire code execution, which is actually not required in practice.

\paragraph{Our Goals.} 
We observe that usually the majority of the code is not security-critical. Typically, a small portion of the code is security-critical and requires cache-based side-channel resilience.
Moreover, this security-critical portion of the code is often already running in an isolated environment, such as in a TEE or in an isolated process.
In these cases, a trusted component, namely the processor hardware or microcode or the operating system kernel, enforces this isolation. We aim to leverage and extend this existing isolation mechanism to also selectively enable side-channel resilience for the caches \emph{only} for the portion of the code that needs it, without reducing the cache performance for the remaining non-isolated code. 
In doing so, we practically address the persistent performance-security trade-off of caches by providing the system administrator with a "tuning knob" to configure by balancing and isolating the workload as required. Consequently, s/he can tune the resulting cache side-channel resilience, utilization, and performance, while guaranteeing no performance overhead is incurred on the non-isolated portion of the code execution.
Only the isolated (usually the minority) portion is subject to a reasonable reduction in cache capacity and performance -- the cost of increased security guarantees.

To achieve this flexible and hybrid cache behavior, we introduce \sys, a generic mechanism that protects isolated code from cache side-channel attacks without reducing the cache performance for the remaining non-isolated code.
In \sys, isolated execution only uses a pre-defined (small) number of cache ways\footnote{\emph{Ways} are different available entries in a cache set to which a particular memory address can be allocated.} in each set of a set-associative cache.
It uses these ways fully-associatively, while for eviction random victim cache lines are selected to be replaced by new ones, thus breaking the set-associativity and removing the root cause of access leakage.
Non-isolated execution uses all cache ways set-associatively as usual, without any performance overhead.
While isolated and non-isolated execution may compete for the use of some ways in the cache, the random replacement policy and fully-associative mapping used by the isolated execution prevent leaking information about the accessed memory locations (and their cache set mapping) to the non-isolated execution, thus making the pre-computation and construction of an eviction set impossible.
Moreover, \sys flexibly supports multiple, mutually distrusting isolated execution domains while preserving the above security guarantees individually for each domain. 

\sys is architecture-agnostic, and can be seamlessly integrated with any isolation mechanism (TEEs or inter-process isolation); the definition of the isolation domains and the distribution of the workload is left up to the system administrator.
\sys is backward compatible by design; it provides conventional set-associative caches for the workload if the side-channel resilience feature is not supported.

\paragraph{Contributions.} The main contributions of this paper are as follows.
\begin{itemize}[noitemsep,topsep=0pt]
	  \item We present \sys, the first cache architecture designed to provide flexible configuration of cache side-channel resilience by selectively enabling it for isolated execution without degrading the performance and available cache capacity of non-isolated execution.
	  \item We evaluate the performance overhead of a simulator-based implementation of \sys and show that it is less than 5\% for the SPEC2006 benchmarks suite, and estimate the memory and area overheads of a cycle-accurate hardware implementation of \sys.
	  \item We show -- through our security analysis -- how breaking set-associative mapping and shared cache lines between mutually distrusting isolation domains (which are the root causes for typical cache side-channel attacks besides the intrinsic cache sharing and competition) mitigates typical contention-based and access-based cache attacks.
\end{itemize}

%% file: background.tex
%!TEX root = main.tex
\section{Cache Organization, Attacks and Defenses} \label{sec:background}
We briefly present the typical cache organization, as well as recent cache side-channel attacks that are within the scope of our work, and limitations of existing defenses.   
\subsection{Cache Organization}
\paragraph{Cache Structure.} Caches are typically arranged in a hierarchy of fastest/closest/smallest to slowest/furthest/largest levels of cache, respectively L1, L2, and L3 cache/last-level-cache (LLC). Each core incorporates its L1 and L2 caches and shares the LLC with other on-chip cores.
A cache consists of the storage of the actual cached data/instructions and the \emph{tag} bits of their corresponding memory addresses. 
Cache memory is organized into fixed-size memory blocks, called \emph{cache lines} each of size \emph{B} bytes. 
Set-associative caches are organized into \emph{S} sets of \emph{W} ways each (called a \emph{W}-way set-associative cache) where each way can be used to store a cache line. A single cache line can only be allocated to only one of the cache sets, but can occupy any of the ways within this cache set.
The least significant $log_2$ \emph{B} bits are the \emph{block offset} bits that indicate which byte block within the \emph{B}-Byte cache line is requested. The next $log_2$ \emph{S} bits are the \emph{index} bits used to locate the correct cache set. The remaining most significant bits are the \emph{tag} bits for each cache line.

In a set-associative cache, once the cache set of a requested address is located, the \emph{tag} bits of the address are matched against the tags of the cache lines in the set to identify if it is a cache hit. If no match is found, then it is a miss at this cache level, and the request is sent down to the next lower-level cache in the hierarchy until the requested cache line is found or fetched from main memory (cache miss). However, in a fully-associative cache, a cache line can be placed in any of the cache ways where the entire cache serves as one set. No index bits are required, but only $log_2$ \emph{B} block offset bits and the rest of the bits serve as tag bits.\\
\vspace{-0.5cm}
\paragraph{Eviction and Replacement.} Due to set-associativity and limited cache capacity, cache contention and capacity misses occur where a cache line must be evicted in favor of the new cache line. Which cache line to evict depends on the replacement policy deployed, some of which include First-in-First-Out (FIFO), Least-Recently-Used (LRU), pseudo-LRU, Least-Frequently-Used (LFU), Not-Recently-Used (NRU), random and pseudo-random replacement policies. In practice, approximations to LRU (pseudo-LRU) and random replacement (pseudo-random) are usually deployed. 

\subsection{Cache Side-Channel Attacks} \label{sec:attacks-rel}
Cache side-channel attacks pose a critical threat to trusted computing and underlie more proliferating side-channel attacks such as the Spectre~\cite{Kocher18} and Meltdown~\cite{Lipp18} variants. Different classes of these attacks have been demonstrated on all platforms and architectures ranging from mobile and embedded devices~\cite{ARMageddon} to server computing systems~\cite{Liu15,Irazoqui15,Zhang12}. 
They have also been shown to undermine the isolation guarantees of trusted execution environments, like Intel SGX~\cite{Brasser17,Schwarz17, Moghimi17, GES17} and ARM TrustZone~\cite{ARMageddon,Zhang16}. Such attacks have been shown to infer both fine-grained and coarse-grained private data and operations, such as \TODO{this is basically the same in intro} bypassing address space layout randomization (ASLR)~\cite{Lipp16, Gras2017}, inferring keystroke behavior~\cite{Gruss16,Gruss15}, or leaking privacy-sensitive human genome indexing computation~\cite{Brasser17}, as well as RSA~\cite{Zhang12,Liu15} and AES~\cite{Bonneau06,Irazoqui15} decryption keys.

Cache side-channel attacks exploit the inherent leakage resulting from the timing latency difference between cache hits and misses.
This is then used to infer privacy/security-critical information about the victim's execution. In an offline phase, the attacker must first identify the target addresses of interest (by means of static and dynamic code analysis of the victim program) whose access patterns leak the desired information about the victim's execution, such as a private encryption key. In an online phase, the attacker measures the timing latency of its memory accesses or the victim's computation time to infer the desired information. 

To demonstrate how a simple cache attack works, consider the pseudo-code of the Montgomery ladder implementation for the modular exponentiation algorithm shown in Algorithm~\ref{fig:algo.rsa}. Modular exponentiation is the operation of raising a number \emph{b} to the exponent \emph{e} modulo \emph{m} to compute $b^e \: mod \: m$ and is used in many encryption algorithms such as RSA. Leaking the exponent \emph{e} may reveal the private key. As shown in Algorithm~\ref{fig:algo.rsa}, the operations performed for each of the exponent bits directly correspond to the value of the bit. If the exponent bit is a zero, the instruction in Line~5 is executed. If the exponent bit is a one, the instruction in Line~9 is executed. An attacker that can observe or deduce these execution patterns can thus disclose the value of each corresponding exponent bit, and eventually recover the encryption key~\cite{Yarom14,Zhang12}. S/he, however, needs to identify the target addresses that need to be observed (the addresses of the instructions in Lines~5 and 9 in this example) in the victim program and accordingly construct the eviction set. The eviction set is a collection of addresses that are mapped to the same specific cache set to which the target addresses are also mapped. The attacker uses this eviction set to evict the contents of the whole set in the cache, and therefore guarantee to successfully evict the target addresses from the caches. Consequently, s/he measures the timing latency of its own memory accesses after a time interval to deduce whether the victim has accessed these target addresses.

\begin{spacing}{1.4}
\RestyleAlgo{ruled}
\begin{algorithm}[h!]{ \footnotesize
\KwIn{base \textit{b}, modulo \textit{m}, exponent \textit{e} = $(e_{n-1}...e_{0})_{\scaleto{2}{3pt}}$}
\KwOut{\textit{$b^e$} $\:$ mod $\:$ \textit{m}}
 $R_0$ $\leftarrow$ 1; $R_1$ $\leftarrow$ b;\\
\For {i from n-1 downto 0}{
	\If {{$e_i$} = 0}{
	$R_1$ $\leftarrow$ $R_0$ $\times$ $R_1$ mod m\;
	$R_0$ $\leftarrow$ $R_0$ $\times$ $R_0$ mod m\;
	}
	\If {{$e_i$} = 1}{
	$R_0$ $\leftarrow$ $R_0$ $\times$ $R_1$ mod m\;
	$R_1$ $\leftarrow$ $R_1$ $\times$ $R_1$ mod m\;
}
}
\Return $R_0$\;
}
\caption{{Montgomery Ladder RSA Implementation} \label{fig:algo.rsa}}
\end{algorithm}
\end{spacing}
The online phase of these attacks consists of three main steps: \emph{Eviction}, \emph{Waiting} and \emph{Analysis}. The attacker uses the eviction set to \emph{evict} the victim's target addresses from the cache. Next, the attacker \emph{waits} an interval of time to allow the victim to access the target addresses. Then the attacker measures and \emph{analyzes} its access time measurements to determine if the victim has accessed the target addresses. This is repeated as many times as the attacker requires to collect sufficient traces to recover the exponent bits. 

The different techniques used by the attacker to perform the eviction can be classified into two main approaches, either access-based or contention-based. In access-based attacks such as Flush + Reload~\cite{Gullasch11,Yarom14}, Flush + Flush~\cite{Gruss16}, Invalidate + Transfer~\cite{Irazoqui16}, and Flush + Prefetch~\cite{Lipp16}, the attacker accesses the target addresses directly by flushing them out of the cache using the dedicated \emph{clflush} instruction~\cite{intel-sw} and possibly exploiting timing leakage from the execution of the \emph{clflush} instruction~\cite{Gruss16}. This invalidates the lines containing these addresses and writes them back to memory. Evict + Reload~\cite{Gruss15} attacks have also been shown which do not require the \emph{clflush} instruction, but instead evict specific cache sets by accessing physically congruent addresses. These attacks are only feasible in case of shared memory pages between the attacker and victim, usually in the form of shared libraries. 
Otherwise, an attacker resorts to contention-based attacks such as Prime + Probe~\cite{Osvik2006,Irazoqui15, Kayaalp16, Liu15, Yan19}, Prime + Abort~\cite{Disselkoen17}, Evict + Time~\cite{Gras2017, Osvik2006}, alias-driven attacks~\cite{Guanciale16}, and indirect Memory Management Unit (MMU)-based cache attacks~\cite{VanSchaik18}, where s/he constructs an eviction set and uses it to trigger and exploit a cache contention in the same cache set as the target addresses, thus evicting cache lines containing the target addresses from the pertinent cache set. 

The waiting interval should be selected and synchronized such that the victim is expected to access the target address at least once before the attacker analyzes the collected observations.
By analyzing the collected observations, the attacker determines whether the target address was indeed accessed by the victim.
This is achieved by different techniques depending on the attack approach, either the adversary measures the overall time needed by the victim process to perform certain computations~\cite{bernstein2005cache, Bonneau06}, or probes the cache with eviction sets and profiles cache activity to deduce which memory addresses were accessed~\cite{Liu15, Irazoqui15, Yan19, Yarom14, Kayaalp16}, or accesses target memory addresses and measures the timing of these individual accesses~\cite{Osvik2006, Gullasch11}.
Alternatively, the adversary can also read values of addresses from the main memory to see whether cache lines that contain cacheable target addresses have been evicted to memory~\cite{Guanciale16}.

Cache-collision timing attacks exploit cache collisions that the victim experiences due to its cache utilization, e.g., after a sequence of lookups performed by a table-driven software implementation of an encryption scheme, such as AES~\cite{Bonneau06}. 
These attacks are out of scope in this work since they are not common, are specific to certain software implementations, and can only be mitigated by adapting the implementation or locking the relevant cache lines after pre-loading them.

\subsection{Limitations of Existing Defenses} \label{sec:defenses}
To mitigate these attacks, software-based countermeasures and modified cache architectures have been proposed in recent years, which we cover in depth in the Related Work (Section~\ref{sec:rel}). These can be classified into two main paradigms: 1) applying cache partitioning to provide strict isolation, or 2) applying randomization or noise to make the attacks computationally impractical.  However, all proposed countermeasures to date either impact performance significantly, require explicit programmer's annotations, are not seamlessly compatible with existing software requirements such as the use of shared libraries, are architecture-specific, or do not defend against all classes of attacks. Most importantly, all existing defenses apply their side-channel cache protection for the entire execution workload. 

In practice, cache side-channel resilience is only required for the security-critical (usually smaller) portion of the workload that is allocated to execute in isolation. Thus, non-isolated execution should not suffer any resulting performance costs.
To address this in this work, we propose a modified hybrid cache microarchitecture that enables side-channel resilience only for the isolated portion of execution, while retaining the conventional cache behavior and performance for the non-isolated execution. 

%% file: attacks.tex
%!TEX root = main.tex
\section{Adversary Model and Assumptions} \label{sec:adv}
To provide side-channel-resilient cache accesses for only security-critical isolated execution, we propose a hybrid \emph{soft} partitioning scheme for set-associative memory structures.
In this work, we apply it to caches and call it \sys. \sys aims to provide cache-based side-channel resilience to the security-critical or privacy-sensitive workload that is allocated to one or more \textbf{I}solated Execution \textbf{D}omains (\sd{}s), while maintaining conventional cache behavior for non-critical execution that is allocated to the \textbf{N}on-\textbf{I}solated Execution \textbf{D}omain (\nsd). \sys assumes an adversary capable of mounting the attacks described in Section~\ref{sec:attacks-rel} and is designed to mitigate them.

Furthermore, the construction of \sys is based on the following assumptions:
\begin{enumerate}[topsep=5.5pt,itemsep=0pt,parsep=0pt,label=\textbf{A\arabic*},ref={A\arabic*},leftmargin=\widthof{\textbf{A1\ \:}}]
\item\label{itm:separated} Security-critical code that requires side-channel resilience is already allocated to an isolated component, like a process or a TEE (enclave).\end{enumerate}
A recent trend in the design of complex applications, like web browsers, is to compartmentalize them using multiple processes.
As an example, all major browsers spawn a dedicated process for every tab~\cite{multiProcessBrowser} and some even use a dedicated process to better isolate privileged components~\cite{chakraJITExternalProcess}.
Similarly, the widespread availability of TEEs, like SGX, encourages developers to encapsulate sensitive components of their code in protected environments.

\begin{enumerate}[resume*]
\item\label{itm:minority} 
Isolated execution is the minority of the workload. 
\end{enumerate}
Isolation works best when the isolated component is as small as possible, thus reducing the attack surface.
This complies with the intended usage of TEEs like SGX where only small sensitive components of the code would be allocated to the TEE. 
Hence, we assume only the minority of the workload needs to be isolated.
\sys still provides the same security guarantees if the majority of the workload is isolated, but the performance of the isolated execution would suffer.

\begin{enumerate}[resume*]
\item\label{itm:shared} Sensitive code only uses writable shared memory for I/O (if at all), and access patterns to this shared memory do not leak any information.
\end{enumerate}
Isolated code should focus on processing some local data, while I/O needs should be limited to copying the input(s) into the isolated component, and copying the output(s) out of the component.
Both of these procedures just access the data sequentially; thus, the access patterns during I/O do not depend on the data and does not leak any information.

\begin{enumerate}[resume*]
\item\label{itm:attacker} The attacker is not in the same \sd as the victim.
\end{enumerate}
\sys is designed to isolate mutually distrusting \sd{}s and thus, we must assume the attacker and the victim are not in the same \sd.
Note that, as a consequence of \ref{itm:shared}, if a process handles sensitive data and has multiple threads, they must all be in the same \sd, since they share the entire address space.
In cases where isolation between threads sharing the same address space is also required, \sys can, in principle, provide intra-process isolation as discussed later in Section~\ref{sec:discuss}.

%% file: design.tex
\section{Hybrid Cache (\sys)} \label{sec:design}
We systematically analyzed existing contention-based and access-based cache attacks in the literature (Section~\ref{sec:attacks-rel}) to identify their common root causes (besides the intrinsic sharing of cache entries and latency difference between a cache hit and miss). 
Cache side-channel attacks are, by nature, very specific to the victim program and may exploit attack-specific features such as the side-channel leakage of the \emph{clflush}~\cite{Gruss16} or prefetch instructions~\cite{Lipp16}. Nevertheless, each one of these attacks is primarily caused by one or both of the following root causes: shared memory pages (and cache lines) between mutually distrusting code, and deterministic and fixed set-associativity of cache structures, which enables targeted cache set contention by pre-computed eviction sets.

\subsection{Requirements Derivation} \label{sec:req}
In light of the above, \sys should provide side-channel resilience between different isolation domains with respect to their cache utilization. 
An adversary process sharing the cache with a victim process should not be able to distinguish which memory locations a victim accesses. 
Nevertheless, we emphasize that the only approach to enforce complete non-interference between different domains is by strict static cache partitioning, such that no cache resources are shared, and thus zero information leakage occurs.
On the other hand, this is impractical, and results in inefficient cache utilization from a performance standpoint.
Our key objective in this work is to practically address and accommodate this persistent performance/security trade-off of cache structures by providing sufficiently strong cache side-channel-resilience, such that practical and typical cache side-channel attacks become effectively infeasible without necessarily enforcing complete non-interference.
Additionally, we desire that this security guarantee is run-time configurable, such that it is only in effect when required.

This builds on our insight that it is neither practical nor required to provide cache side-channel resilience for all the code in the workload. This additional security guarantee is only required for security-critical execution, which is a minority of the workload (Assumption~\ref{itm:minority}), and usually isolated in a Trusted Execution Environment (TEE) (Assumption~\ref{itm:separated}). Thus, we require to provide a cache architecture that provides non-isolated execution with conventional cache utilization (with no performance costs), and simultaneously side-channel-resilient cache utilization (with a tolerable performance degradation) only for the smaller portion of the execution workload that is security-sensitive and isolated. We also require that our architecture is portable, can be easily deployed, and is backward compatible when a system does not support it. We summarize these requirements below:
\begin{enumerate}[topsep=0pt,itemsep=0pt,parsep=0pt,label=\textbf{R\arabic*},ref={R\arabic*},leftmargin=\widthof{\textbf{R1\ \:}}]
\item\label{itm:scresistance} Strong side-channel resilience guarantees between the isolated and non-isolated execution domains, sufficient to thwart typical contention-based and access-based cache attacks
\item\label{itm:cacheisolation} Dynamic and scalable cache isolation between multiple different isolation domains 
\item\label{itm:performance} Addressing the cache performance/security trade-off by configuring the non-isolated/isolated workload balance (compliant with how TEEs are intended and designed to be used) such that the performance of the non-isolated execution workload is not degraded
\item\label{itm:usability} Usability: backward-compatible, architecture-agnostic, no usage restrictions and no code modifications required
\end{enumerate}
Next, we present the high-level construction of \sys in Section~\ref{sec:high-level} and its microarchitecture in more detail in Sections~\ref{sec:controller} and \ref{sec:hw-cache}.

\subsection{High-Level Idea} \label{sec:high-level}
In \sys, a subset of the cache, named \subcache, is reserved to form an orthogonal isolated cache structure.
Specifically, \nsecure \emph{cache ways} within the conventional \emph{cache sets} form the \subcache.
While these \subcache ways are available for the \nsd to utilize, the \sds are restricted to utilize \emph{only} these \subcache ways. However, the \sds utilize this \subcache in a fully-associative way and using a random-replacement policy. In doing so, all mutually distrusting processes executing in the \sds can share the \subcache without leaking information on the actual memory locations they access.
Since these \subcache ways are not reserved exclusively for isolated execution and can also be utilized by non-isolated execution with least priority, the \nsd still retains unaltered cache capacity usage and non-degraded performance.

The key purpose of \sys, unlike existing defenses, is to selectively enable side-channel-resilient cache utilization only for the \sds. Hence, only the isolated execution is subjected to the resulting performance overhead, while still maintaining conventional cache behavior and performance for the \nsd, as outlined in Requirement~\ref{itm:performance}.  We describe next the architecture of \sys and how it achieves this.

\subsection{Controller Algorithm} \label{sec:controller}
\begin{figure*}[t]
    \centering
    \includegraphics[width=1.0\textwidth]{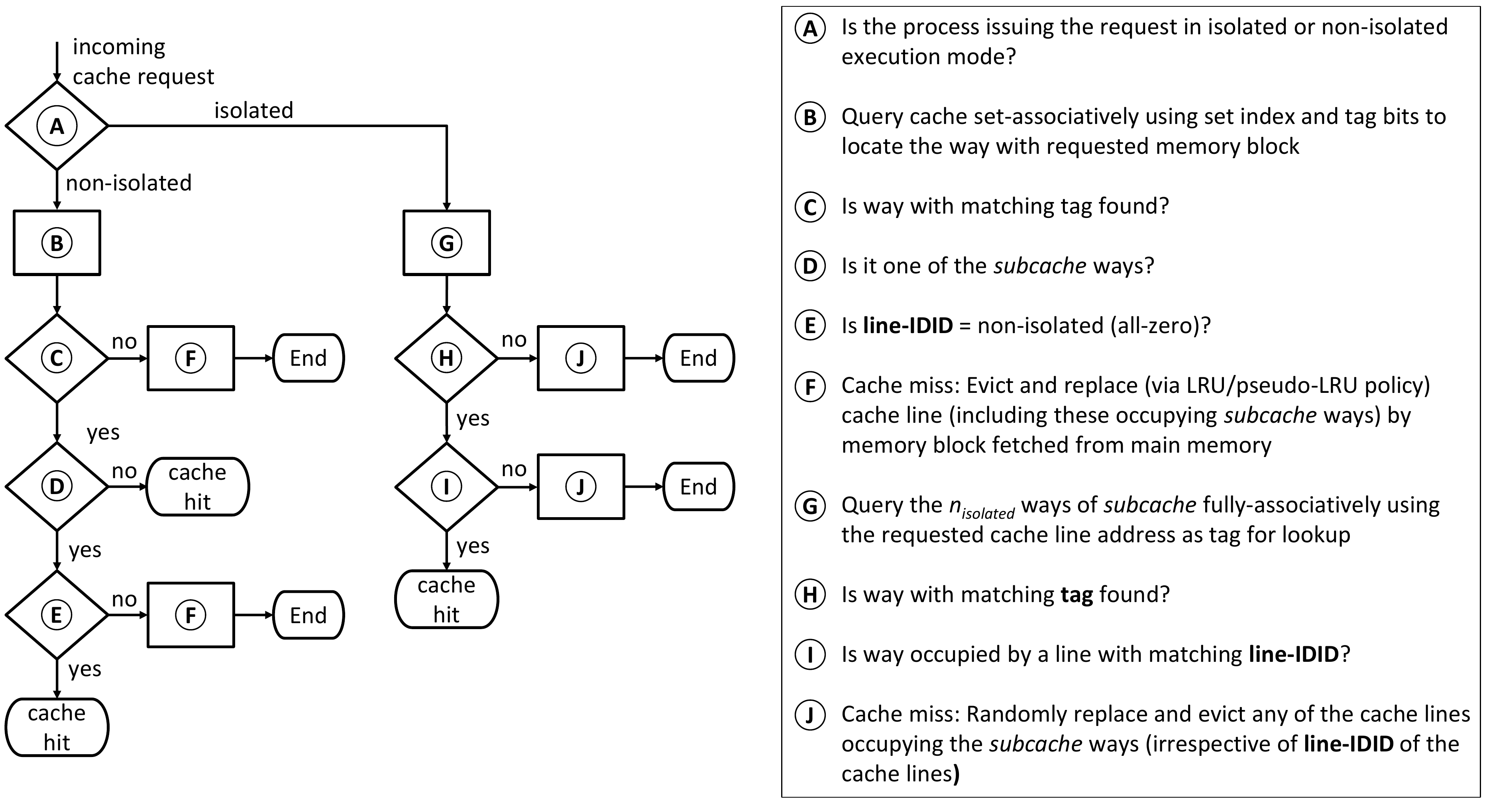}
    \caption{\sys controller policy}
    \label{fig:cache-scheme}
\end{figure*}

\sys modifies how memory lines are mapped to cache entries for the \sds. 
\nsecure ways (at least a way in each set) of the conventional set-associative cache are designated to the orthogonal \subcache. Cache lines are mapped fully-associatively to the \subcache entries and evicted and replaced in the \subcache using a random replacement policy. 
This means that a given memory line can be cached in any of the \nsecure entries. 
This breaks the deterministic link between memory addresses and their corresponding cache locations, thus defeating an attacker that attempts to infer the victim's memory accesses by triggering and observing contention in a particular cache set. 

Figure~\ref{fig:cache-scheme} illustrates how the \sys controller manages cache requests. \sys supports multi-core processors with simultaneous multithreading (SMT) and assumes that each process is assigned an \ididlong (\idid) that identifies whether the process is in an \sd (and which isolation domain) or in the \nsd. Any incoming cache request is accompanied by the \idid of the issuing process. 
In \circled{A}, \sys controller queries the \idid of the cache request and the request is serviced accordingly. If it is in the \nsd, the complete cache is queried conventionally using the set index and tag bits of the requested address to locate the cache set and line respectively (\circled{B} \& \circled{C}). 
If a match is found, the controller checks whether the cache line was found in one of the \subcache ways in \circled{D}. Recall that these ways are not reserved exclusively for isolated execution, i.e., they can be used by non-isolated execution but with least priority in case a cache set becomes over-utilized. Therefore, if a matching cache line is found in one of these ways, the controller checks whether it was cached by an isolated or non-isolated process (\circled{E}). 
The requesting process can only hit and access the cache line if that line was placed by a process in the \nsd. Otherwise, it is not allowed to hit on it.

Checks in the controller are implemented to occur in parallel, i.e., all cache hits are generated in the same number of clock cycles (as well as cache misses), to eliminate respective timing side channels.
In case of a cache miss, the memory block is fetched from main memory and cached in \circled{F}. The eviction and replacement are performed according to the deployed policy.
All ways are available for eviction, including the \subcache ways to provide the \nsd execution with unaltered cache capacity. However, the usage of the \subcache ways by the \sds is considered while recording the recency of accesses to the cache ways to make it least likely to evict a line from one of the \subcache ways if it is recently used by an \sd process.

If the cache request is issued by an \sd process, it is serviced by querying only the \subcache (\circled{G}). The \subcache deploys fully-associative mapping, and is thus queried by a lookup of all the ways using the (cache line address bits - block offset bits) as tag bits (\circled{H}) and simultaneously querying that the line belongs to an \sd (since these ways may also be used by the \nsd) and that it was placed by a process with the same \idid  (\circled{I}). 
Otherwise, a cache miss occurs.
Disallowing \sd processes from hitting on cache lines originally placed by processes in other \sds provides dynamic isolation between an unlimited number of mutually distrusting processes that share memory. In case of a miss, any of the \subcache ways is randomly selected and its cache line is evicted and replaced by the memory block fetched from main memory (\circled{J}). The random replacement policy considers all \subcache ways equally, even those occupied by the \nsd cache lines.

\subsection{Hardware Microarchitecture} \label{sec:hw-cache}
\begin{figure}[!ht]
    \centering
    \includegraphics[width=1.0\columnwidth]{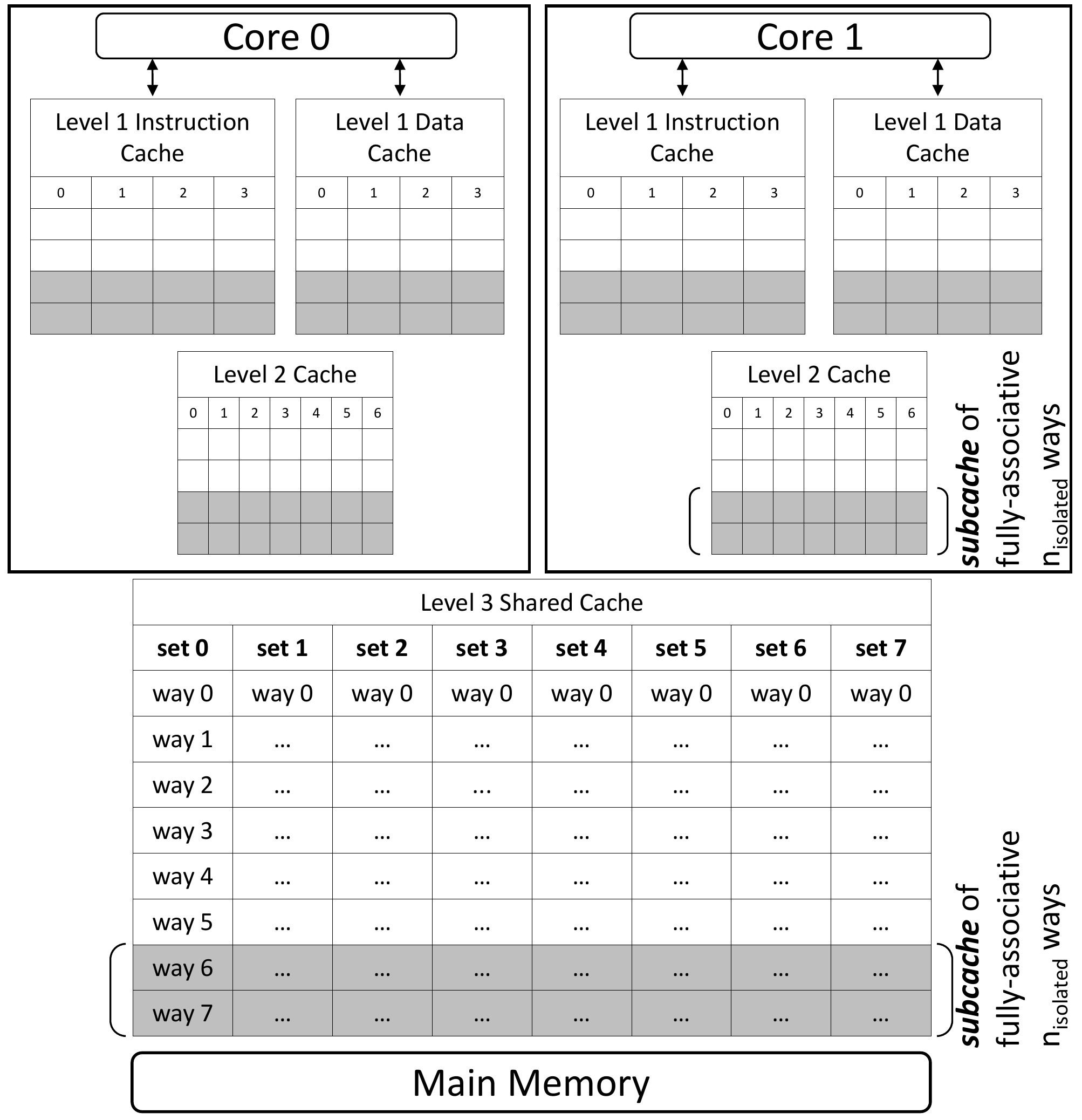}
    \caption{\sys hierarchy and organization}
    \label{fig:hw-cache}
\end{figure}

Figure~\ref{fig:hw-cache} shows how \sys could be applied for a conventional cache hierarchy of a multi-core processor. The cache capacity available for the \nsd execution is unaltered, i.e., the conventional set-associative cache with all its sets and ways can be utilized by the \nsd.

At each cache level, way-based partitioning is used to reserve at least a way in each set (gray ways in Figure~\ref{fig:hw-cache}). 
These ways, combined, form the orthogonal \subcache that the \sd execution is restricted to use.   
However, these \subcache ways are \emph{not} used exclusively by the \sd execution, i.e., the \nsd execution may use these ways in case a corresponding set is fully utilized and the least-recently-used (LRU) replacement algorithm requires to evict a cache line from a \subcache way in this set. This ensures that the \nsd execution is provided with unaltered cache capacity and does not suffer performance degradation.

The \subcache is fully-associative and deploys random replacement policy, i.e., a given memory block is always equally likely to be cached in any of the available ways.
This breaks set-associativity and provides randomization-based dynamic isolation between different \sds while allowing flexible sharing of the \subcache depending on the run-time utilization requirements of the isolated execution domains. Using the \subcache fully-associatively further maximizes the utilization of its limited hardwired capacity. 

\vspace{1em} % orphan avoidance

\begin{figure}[h]
    \centering
    \includegraphics[width=1.0\columnwidth]{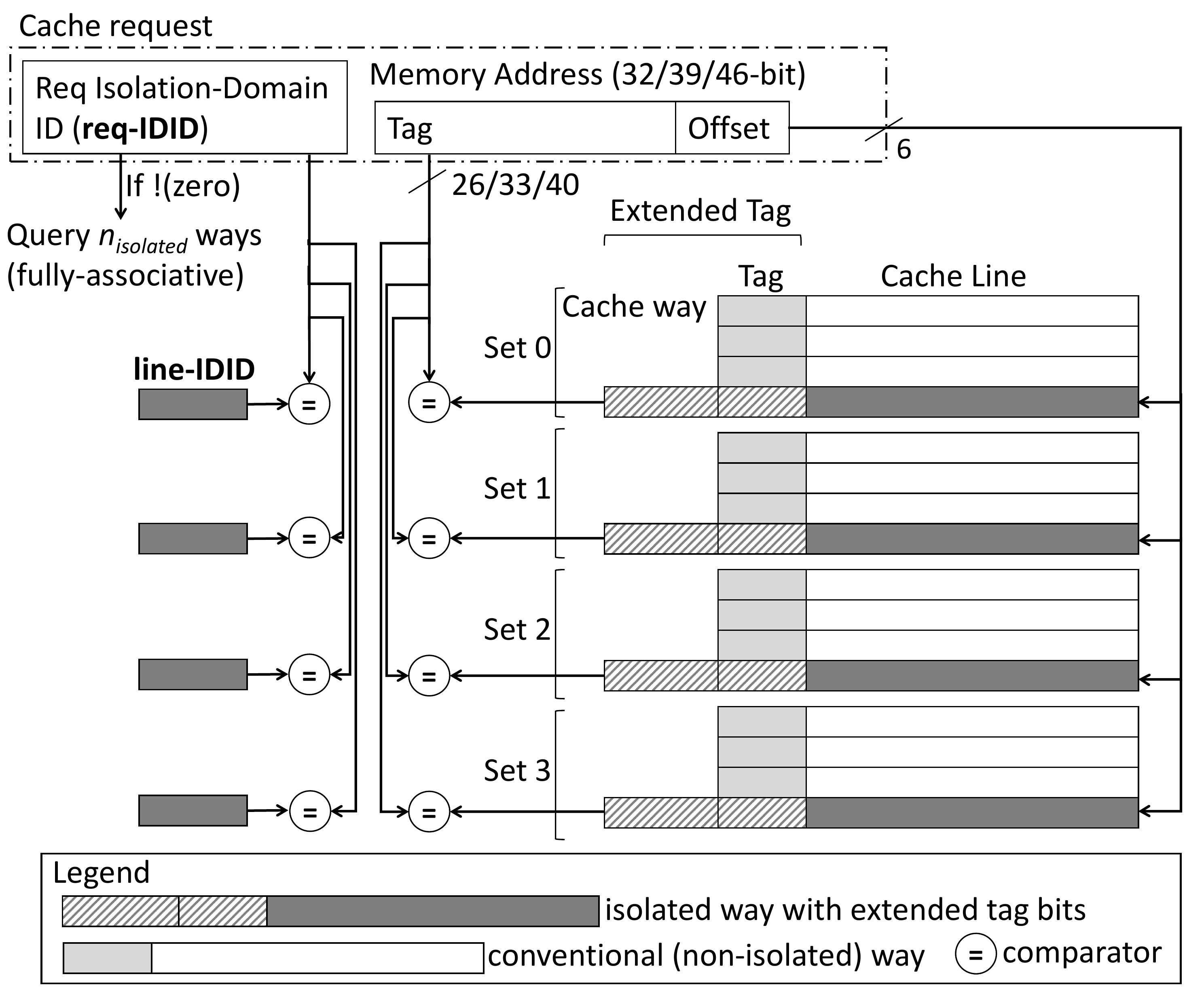}
    \caption{\sys hardware microarchitecture}
    \label{fig:sec-ways}
\end{figure}

The \nsecure ways that form the \subcache are configured (hardwired) at design-time and cannot change at run-time, because these ways are members of both the primary cache as well as the \subcache as shown in Figure~\ref{fig:sec-ways}. It is not feasible to make \nsecure run-time configurable, as this would require that \emph{all} the ways are unreasonably wired in both a fully-associative and set-associative organization. Thus, only a small subset of \nsecure ways (dark gray ways in Figure~\ref{fig:sec-ways}) is selected to form the \subcache. 
Each of the \subcache ways is augmented with \ididlong (\idid) configuration bits to identify the isolation domain that placed an occupying cache line in the pertinent way.
To provide any cache isolation at the microarchitectural level, a mechanism to bind owners/tags to cache lines is required, thus \idids are needed.  We chose to configure 4 bits for the \idid, thus supporting 16 concurrent isolation domains, where an all-zero indicates the \nsd. The number of bits allocated in \sys for \idid is a hardware design decision. Increasing the number of designated bits would increase the number of maximum concurrent isolation domains that \sys can support. However, other metrics such as area overhead and power consumption come into play in this design trade-off.

The \subcache ways are augmented with an extended tag bits storage (dashed dark gray tag bits of the dark gray ways in Figure~\ref{fig:sec-ways}). When queried fully-associatively (for the \sds), all bits, except the offset bits (6 bits for byte-addressable 64B cache line), of the requested address are compared with the extended tag bits of the \subcache ways to locate a matching cache line. For the \nsd, the \subcache ways are queried set-associatively with the rest of the cache (conventionally), where the request tag bits are compared only with the non-extended tag bits of the \subcache ways within the located cache set.

\subsection{Software Configuration} \label{sec:sw-cache}
\paragraph{Abstraction and Transparency.}  The hardware modifications required for \sys are transparent to the software and abstracted from it. The trusted software (or hardware) component of the incorporating platform is only required to interface with the \sys controller to communicate the isolation domain of each incoming cache request. However, \sys does not stipulate or restrict how these isolation domains are defined and communicated, thus leaving it to the discretion of the system designer to identify how \sys can be integrated with the comprising architecture. 

\paragraph{Isolated Execution.} \sys enables the dynamic isolation of the cache utilization of different isolation domains by using the \idid of the process that issues the cache request being serviced. The means by which the isolation domains are defined, generated, and communicated is dependent on how the trusted execution and isolation is deployed.
We design \sys such that it is seamlessly compliant with any trusted execution environment (TEE) where isolation domains (across different processes, cores, containers, or virtual machines (VMs)) are either software-defined by a trusted OS (thus requiring kernel support) or hardware/firmware-defined in case the OS is not trusted (such as in SGX). 
Different isolation domains can be defined across different isolated address space ranges such as in SGX enclaves, across processes such as in TrustZone normal/secure worlds or by standard inter-process isolation, or even across different groups of processes or different virtual machines.

\sys is agnostic to the means of defining the \idids of  different isolation domains, and complements any form of isolated execution environment in place to provide it with cache side-channel resilience.
If the kernel is trusted, kernel support is required to assign an \idid (or an all-zero \idid for a non-isolated process) to each process according to its isolation domain.
The \idid bits can be added as an additional process attribute in each process's process control block (PCB).
Otherwise, the trusted hardware or firmware would assign the isolation domains.
\sys assumes that some mechanism of isolation is already enforced for security-critical code that it can leverage to provide the cache-level isolation. We argue why this is reasonable in Assumption~\ref{itm:separated}. Nevertheless, if this is not the case, then isolation domains need to be explicitly defined by the developer if s/he wishes to protect particular code against cache-based side-channel attacks. While \sys is focused on protecting user code, in principle, kernel code can also be protected by allocating it to an isolation domain.

\paragraph{Backward Compatibility.} Similar to processor supplementary capabilities such as Page Attribute Tables (PATs) and Memory Type Range Register (MTRR) for x86, \sys supports providing side-channel-resilience on-demand while retaining backward compatibility. \sys only effectively provides side-channel resilience for the cache utilization of execution when processes are assigned different \idids that are communicated with each cache request. Otherwise, from a software perspective, \sys is identical to a conventional cache architecture. 
If no isolation domains are assigned to the different processes by the trusted kernel or trusted hardware, \sys is designed to assign an all-zero \idid by default to incoming cache requests and all execution is treated as non-isolated (see Figure~\ref{fig:cache-scheme}) with cache-based side-channel resilience disabled. 
Only when kernel support is provided (or trusted hardware or firmware in case of SGX) %to assign an \idid to each process does 
does \sys behave differently for different isolation domains and provides its side-channel resilience capability.

\paragraph{Shared Memory Support.} \sys supports, by design, that different isolation domains can share read-only memory, usually in the form of shared code libraries, without sharing the corresponding cache lines. This results in having multiple copies of the shared memory kept in cache (multiple cache entries), enforcing that cache entries are not shared between mutually distrusting code. Data coherence is also not a problem, in this case, since this is read-only memory. We elaborate in Section~\ref{sec:security} how this effectively mitigates access-based side-channel attacks.

Conventional access to shared writable memory, on the other hand, between different isolation domains is disallowed by design in \sys, as this makes the victim process vulnerable to access-based attacks and would undermine cache coherence.
In order to provide input and output functionality to isolated code, \sys provides special \emph{I/O move instructions}. These allow code in an \sd to transfer data between a CPU register and a memory region (assigned an all-zero \idid when cached) that is designated exclusively for shared memory between processes belonging to different \sds.
These special instructions are meant to be used to transfer data between domains only through this designated memory. In practice, we expect them to be used only in frameworks like the SGX SDK or a trusted kernel.
If code in an \sd incorrectly accesses this memory region using regular instructions, or accesses its own memory using these special instructions, this could be disallowed, i.e., detected and blocked by the hardware or microcode, e.g., the MMU. This prevents inserting duplicated writable cache entries which can disrupt cache coherency, while ensuring that \sys's security guarantees still apply to any access performed using regular instructions.

%% file: security.tex
%!TEX root = main.tex
\section{Security Analysis} \label{sec:security}
In the following, we evaluate the effectiveness of \sys with respect to the security requirements we outlined in Section~\ref{sec:req}. We show that \sys achieves these security guarantees by mitigating the following leakages:

\begin{enumerate}[topsep=0pt,itemsep=0pt,parsep=0pt,label=\textbf{S\arabic*},ref={S\arabic*},leftmargin=\widthof{\textbf{S1\ \:}}]
\item\label{itm:nodirectcheck} Malicious software running in an \sd or \nsd cannot flush or perform a cache hit on a cache line belonging to a different \sd.
\item\label{itm:noevictionset} Malicious software running in an \sd or \nsd cannot pre-compute and construct an eviction set that selectively evicts a non-trivial subset of the cache lines belonging to a different \sd. Moreover, the set of the attacker's cache lines which can be evicted by the victim's lines does not depend on the addresses accessed by the victim. 
\item\label{itm:missstillleak} Cache hits generated by software in an \sd cannot be observed by software running in a different \sd or \nsd. Cache misses generated by software in an \sd can still be indirectly observed by malicious software running in a different \sd or \nsd, but the malicious software learns no information (e.g., memory address) about the access besides whether a cache miss has occurred.
\end{enumerate}

\subsection{\ref*{itm:nodirectcheck}: Absence of Direct Access to Cache Lines}
Access-based attacks, like Flush + Reload~\cite{Gullasch11,Yarom14}, Flush + Flush~\cite{Gruss16}, Invalidate + Transfer~\cite{Irazoqui16}, Flush + Prefetch~\cite{Lipp16}, and Evict + Reload~\cite{Gruss15}, require the attacker to have direct access to the victim's cache lines, normally as a result of shared memory between processes (e.g., shared libraries).
As an example, Flush + Reload works by flushing shared cache lines and monitoring which lines the victim accesses and brings back into the cache.
\sys mitigates this class of attacks by preventing shared cache lines between the attacker and victim, as we explain in the following.

\paragraph{Shared Read-Only Memory.}
Read-only memory is shared between different processes in case of shared code libraries. \sys provides support for shared read-only memory (Section~\ref{sec:sw-cache}), while fundamentally disallowing that any cache line is shared across different \sd{}s.
Execution within one domain can only access cache lines brought into the cache by the same domain.
Separate (potentially duplicate) cache lines are maintained for each domain; flushing and reloading cache lines only impacts those owned by the attacker's domain and cannot influence any other \sd or leak any information on its cache lines.
Having duplicate cache lines for read-only memory pages does not disturb cache coherency because it is read-only.

\paragraph{Shared Writable Memory.}
Shared writable memory between mutually distrusting domains is disallowed by design with \sys. %, which is a reasonable caveat.
Code in an \sd can only exchange data with another isolation domain through the special I/O move instructions, which transfer data between the CPU registers and memory in the \nsd that is designated for shared communication (see Section~\ref{sec:sw-cache}).
Incorrect usage of those instructions or incorrect access to this designated memory region could be detected and blocked by the MMU to prevent potential cache coherency disruption due to duplicate writable cache entries. However, \sys still enforces that every cache line only belongs to one domain. Since cache lines always belong to one specific \sd or the \nsd, code in a domain cannot flush or perform a cache hit on a different domain's cache lines (\ref{itm:nodirectcheck}), and attacks that rely on those capabilities are thus impossible.

\subsection{\ref*{itm:noevictionset}: Impossibility of Pre-Computed Eviction Set Construction}
Without direct access to the victim's cache lines, attackers resort to contention-based attacks, like Prime + Probe~\cite{Osvik2006,Irazoqui15, Kayaalp16, Liu15, Yan19}, Prime + Abort~\cite{Disselkoen17}, and Evict + Time~\cite{Gras2017, Osvik2006}.
In these attacks, the attacker pre-computes and constructs an eviction set which ensures eviction of a specific subset of the victim's cache lines, e.g., lines that belong to a specific set in a set-associative cache.
The attacker process first accesses the whole eviction set, thus ensuring the victim's cache lines are evicted. After a waiting interval, it then checks if its whole eviction set is still in cache by timing its own memory accesses to this set, thus detecting if the victim accessed any of the cache lines of interest.
For a conventional set-associative cache, this is possible because of a fixed set-indexing, which can be directly determined from the target address of interest.

\sys protects \sd{}s from such attacks by disabling the set-associativity of the reserved \subcache entries when they are used by isolated execution: when a memory address is accessed by the isolated victim process, the cache line will be stored in any entry chosen randomly from the whole \subcache and not from a specific set.
The random replacement policy for isolated execution ensures that any of the \subcache entries is chosen using a discrete uniform distribution, i.e., with an equal and independent probability every time, so
the attacker has no means of identifying deterministically and reproducibly which cache set (or entry) will be used to cache a particular memory access of the victim.
In order to ensure that a specific cache line of the victim is evicted, the attacker can only evict all lines in the \subcache, but s/he cannot selectively evict a non-trivial subset of the victim's cache lines.
Moreover, the set of the attacker's cache lines which can be evicted by the victim's lines does not depend on the addresses accessed by the victim (\ref{itm:noevictionset}).
As a consequence, attacks that rely on these capabilities are no longer possible.
This holds whether the attacker process is running in an \sd or \nsd, as long as the victim process is in an \sd (Requirements \ref{itm:scresistance} and \ref{itm:cacheisolation}).

\subsection{\ref*{itm:missstillleak}: Observable Cache Events} \label{sec:occupancy}

Software running in an \sd can only hit on cache lines belonging to the same \sd. These cache hits generate no changes to the cache state, thus, they are unobservable by an attacker in a different \sd or in the \nsd.

Cache misses generated by software in an \sd evict a random cache line, which may belong to a different \sd or  the \nsd.
Malicious attacker code can then periodically observe how many of its lines are evicted and infer the number of cache misses the victim process is experiencing.
The attacker can further use this information to infer the size of the victim's working set, i.e., the number of cache lines in the \subcache currently belonging to the victim.

This cache occupancy channel is the only side-channel leakage that is not mitigated by the \sys construction, which is inherently available in any cache architecture where the attacker and the victim processes compete for entries in shared cache resources.
It can only be effectively blocked by strict cache partitioning, which we deliberately do not provide in the \sys construction. This allows different isolation domains to still compete for cache entries, thus preserving maximum and dynamic cache utilization and unaffected performance for non-isolated execution, as our performance evaluation shows in Section~\ref{sec:perfeval}. Note that, due to \ref{itm:noevictionset}, the information inferred by the attacker from observing this remaining leakage, is effectively reduced to only knowing the working set size at any point in time.

Leveraging this side channel to infer further information and mount an attack in typical settings is not trivial.
The victim may evict its own lines when it experiences cache misses due to the random replacement policy. This would not effect a difference in the cache state for the attacker, which complicates the attacker's bookkeeping.
Moreover, observations are severely hindered when any other software is concurrently running besides the attacker and the victim processes.
Finally, standard software hardening techniques can be applied to mitigate attacks to code implementations that are particularly sensitive to this attack.
Furthermore, exploiting this side channel to leak data has not been shown in practice.
A recent attack~\cite{WebFingerprinting} leverages the cache occupancy side channel to infer which website is open in a different browser tab (under the strong assumption that no other tabs are open); however, it does not leak any user data. Cache activity masking is suggested as one of the countermeasures to the attack.
Implementing cache activity masking for \sys is feasible and independent of our cache architecture.

Since the attacker aims to maximize its information and cannot observe cache hits, s/he can attempt to evict all \subcache entries in order to maximize the number of misses experienced by the victim.
As we discuss later, evicting the whole \subcache takes time for an attacker in either the \nsd or in a \sd.
An unprivileged attacker is unable to pause the victim's execution; thus, the attacker can only measure the cache usage with limited granularity.
However, a privileged adversary, like a malicious OS in the case of an SGX enclave, can stop and restart the victim arbitrarily and leverage tools like SGX-Step~\cite{sgxstep} to observe the victim's cache usage with fine granularity. \sys does not mitigate such an attack by construction. However, mitigating it is only possible by strict cache partitioning and the resulting performance costs. We emphasize that we make an intentional design decision in \sys to allow isolation domains to dynamically compete for cache entries for maximum cache utilization and unaffected performance for non-isolated execution. A \sys construction that dynamically allocates a dedicated \subcache for each isolation domain would block this leakage and mitigate attacks that rely on it.

\paragraph{Non-isolated Attacker Process.}
If the attacker process is in the \nsd, in order to guarantee eviction of the whole \subcache it must fill up all ways in every cache set, including the \subcache ways.
Therefore, the attacker process must construct an eviction set that is as large as the entire cache capacity.
A typical data L1 cache holds 512 cache entries.
In our experiments, probing (accessing and measuring access latencies) of 512 cache lines takes approximately 30\,000 CPU cycles, i.e.,  a little over 8~\textmu{}s.\footnote{We ran this experiment on an Intel i7-4790 CPU clocked at 3.60~GHz.}
For larger caches, such as the LLC, it is not even feasible to mount Prime+Probe attacks by probing the entire cache. The adversary is required to pinpoint a few cache sets that correspond to the relevant security-critical accesses made by the victim and monitor these only~\cite{Liu15}. 

\paragraph{Isolated Attacker Process.}
If the adversary is in a different \sd than the victim process, it still cannot control cache eviction of particular target addresses specifically. 
Both attacker and victim processes are isolated and can only use the \subcache ways. 
Thus, an adversary aiming to perform controlled eviction can only try to evict the entire \subcache. Because the \subcache is fully-associative with random replacement, evicting the entire \subcache requires an eviction set much larger than the \subcache capacity.
We argue below that this is not easier than probing the entire L1 cache (in case the attacker is non-isolated), for instance, even though the \subcache is significantly smaller.
Moreover, it can be only guaranteed up to a certain level of probabilistic confidence.
This can be represented statistically by the coupon collector's problem, where coupons are represented by entries in the \subcache.
Let $N_{accesses}$ be the total number of accesses needed to evict all the \subcache entries $n$ and $n_i$ be the number of accesses needed to evict the $i$-th way after $i$-$1$ ways have been evicted.
Both $N_{accesses}$ and $n_i$ are discrete random variables. 
The probability of evicting a new way becomes $\frac{(n - (i-1))}{n}$.
The expected value and variance of $N_{accesses}$ are 
\begin{equation*}
\mathbb{E}(N_{accesses}) = n\cdot\mathrm{H}_{n} \qquad
\mathbb{V}(N_{accesses}) \approx \frac{\pi^2}{6}\cdot n^2
\end{equation*}

\noindent $\mathrm{H}_{n}$ denotes the $n^\text{th}$ harmonic number.
For $n$ = 128 \subcache entries, an average of 695 memory accesses (each mapping to a different 64B cache line) is needed to evict the \subcache with a variance of $\approx$ 26\,951. This is comparably more than the 512 accesses required to probe the entire typical L1 cache if the attacker process is not isolated (see above).
Moreover, with such a large variance, significant variations in the number of $N_{accesses}$ required are expected from the mean $\mathbb{E}(N_{accesses})$ every time this eviction process is repeated.

%% file: evaluation.tex
%!TEX root = main.tex
\section{Evaluation} \label{sec:eval}
\begin{table}[hb]
	\centering \small
	\begin{tabular}{crrr}
	\toprule
	\bfseries Cache & \multicolumn{1}{c}{\bfseries Size} & \multicolumn{1}{c}{\bfseries Associativity} & \multicolumn{1}{c}{\bfseries Sets} \\
	\midrule
	L1 & 64~KB & 8-way associative & 128 \\
	L2 & 256~KB & 8-way associative & 512 \\
	L3 & 4~MB & 16-way associative & 4096 \\
	\bottomrule
	\end{tabular}
	\caption{Cache hierarchy used in our evaluation}
	\label{tab:caches}
\end{table}

\begin{table}[hb] % move this table after the other if the other doesn't fit on top of the graphs
	\centering
	\begin{tabular}{>{\ttfamily}ccc}
	\toprule
	\rmfamily \bfseries Mix & \rmfamily \bfseries Components \\
	\midrule
	pov+mcf & povray, mcf        \\
	lib+sje & libquantum, sjeng  \\
	gob+mcf & gobmk, mcf         \\
	ast+pov & astar, povray      \\
	h26+gob & h264ref, gobmk     \\
	bzi+sje & bzip2, sjeng       \\
	h26+per & h264ref, perlbench \\
	cal+gob & calculix, gobmk    \\
	\midrule
	pov+mcf+h26+gob & povray, mcf, h264ref, gobmk \\
	lib+sje+gob+mcf & libquantum, sjeng, gobmk, mcf \\
	\bottomrule
	\end{tabular}
	\caption{Benchmark mixes used in our evaluation}
	\label{tab:procpairs}
\end{table}

\sys is architecture-agnostic and applicable to x86, ARM or RISC-V. We performed our performance evaluation of \sys on a gem5-based~\cite{gem5} x86 emulator. We evaluated the hardware overhead for an RTL implementation that we implemented to extend an open-source RISC-V processor Ariane~\cite{ariane}. For our prototyping, we applied \sys to L1, L2, and LLC. We describe our evaluation results next.

\subsection{Performance Evaluation} \label{sec:perfeval}
To evaluate \sys, we chose eight mixes of programs from the SPEC CPU2006 benchmark suite, which are used in the literature\footnote{\cite{Yan17} also uses a ninth mix, \texttt{dea+pov}, which fails to run on gem5.}~\cite{Yan17,Jaleel10}, shown in the upper part of Table~\ref{tab:procpairs}.
 
\paragraph{Two-Process Mixes.} In order to evaluate the impact of isolating one process in the context of an SMT processor, we configure gem5 to simulate two processors connected to a single three-level cache hierarchy, whose parameters are shown in Table~\ref{tab:caches}. The caches have the latencies used in~\cite{Yan17}.

For each mix, we first isolate one process, then the other, and we compare the performance of those processes to a third run in which neither process is isolated.
We make either 2 or 3 of ways per set usable by the isolated execution processes. The replacement policy for non-isolated processes is LRU.
Like in~\cite{Yan17}, we let gem5 simulate the first 10 billion instructions of each process in order to let the process initialize, then we measure the performance of one additional billion instructions.
We measure the performance overhead as the relative change in the instructions-per-cycle (IPC), i.e., the ratio between instructions executed and CPU cycles required.
A \emph{positive} overhead represents a \emph{decrease} in performance.

\begin{figure}[p]
	\centering
	\includegraphics[width=\columnwidth]{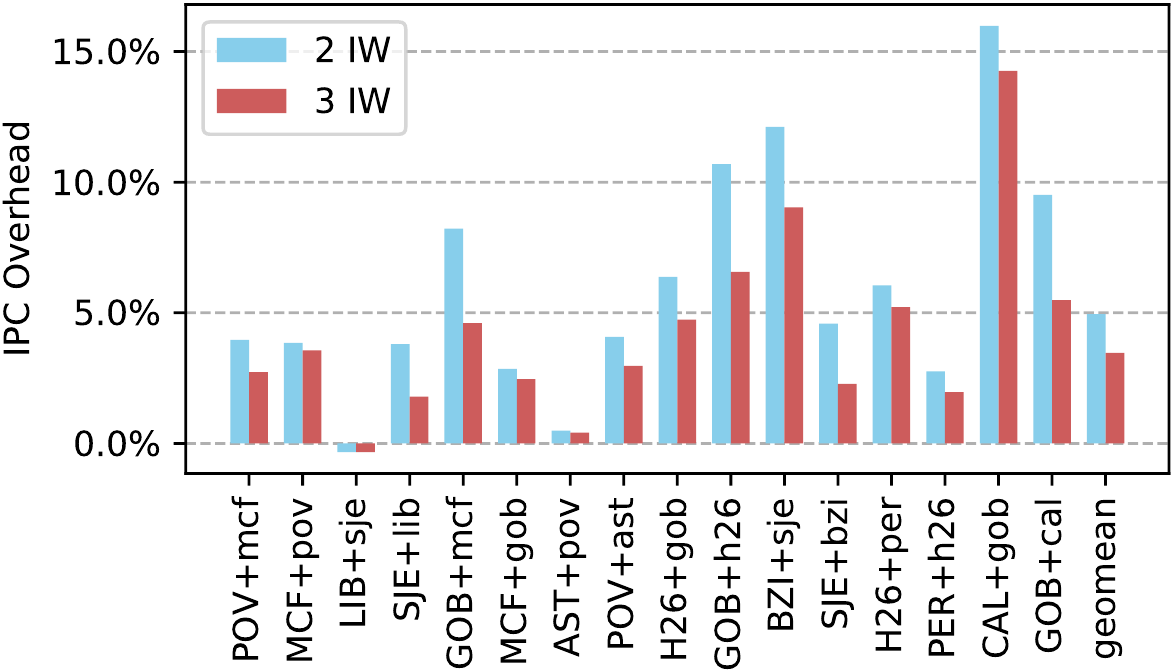}
	\caption{IPC overhead of each isolated process when 2 or 3 ways are available to isolated execution. Each pair of bars refers to a specific 2-process mix: the uppercase benchmark is isolated and the other is not.}
	\label{fig:2ovh-iso}
\end{figure}

\begin{figure}[p]
	\centering
	\includegraphics[width=\columnwidth]{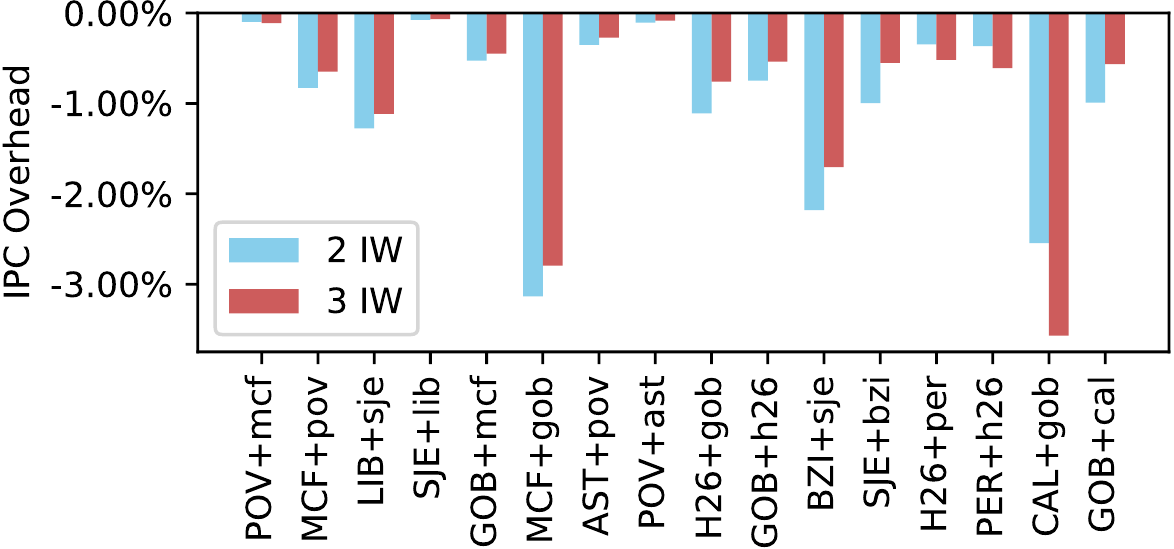}
	\caption{IPC overhead of each process when the other member of the mix is isolated. Each pair of bars refers to a specific 2-process mix: the uppercase benchmark is isolated and the other is not.}
	\label{fig:2ovh-niso}
\end{figure}

Figure~\ref{fig:2ovh-iso} reports the IPC overhead of each program when running in isolation mode, while the other member of the mix runs in normal mode, for 2~or 3~isolated ways.
The geometric mean of the positive overheads is 4.95\% with 2~isolated ways and 3.47\% with 3~isolated ways, with maximum overheads of 16\% and 14\% respectively for the \texttt{cal+gob} mix.
For this mix, the overhead is due to a significantly increased L3 cache miss rate: the data miss rate jumps from 0.6\% to 17.6\%, while the instruction miss rate increases from 2.1\% to 9.0\%.
The working set of \texttt{calculix} normally fits in L3~\cite{Jaleel10} but it does not in the \subcache, hence the higher overhead.
Since \sys is meant to protect only sensitive applications, which can be expected to be short-lived and only constitute a minority of the workload of a system, we consider those overheads easily tolerable.
Figure~\ref{fig:2ovh-niso} reports the IPC overhead for the member of the mix that is not isolated.
In all cases the IPC overhead is not positive, i.e., the IPC is equal or better than the baseline, thus showing that \sys does not degrade the performance of non-isolated processes.

\begin{figure}[p]
	\centering
	\includegraphics[width=\columnwidth]{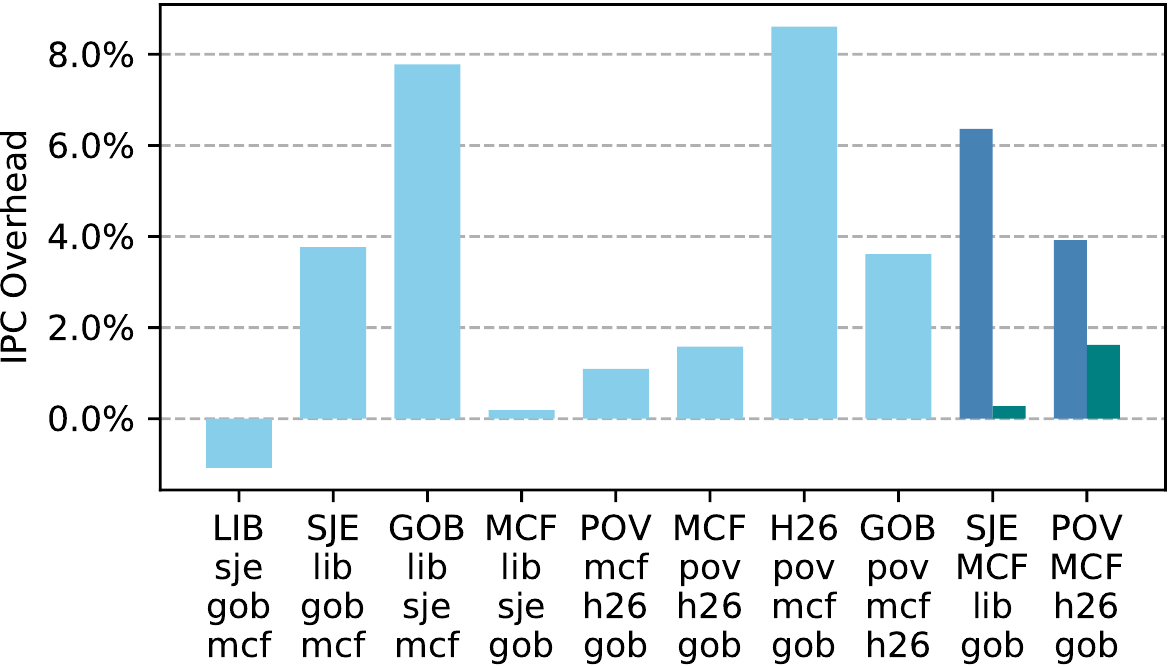}
	\caption{IPC overhead of isolated processes for 4-process mixes. The uppercase benchmarks are isolated and the others are not. The last two columns have two bars each since two process are isolated.}
	\label{fig:2ovh-4}
\end{figure}

\paragraph{Four-Process Mixes.} To demonstrate scalability, we also ran four-process mixes, shown in the bottom part of Table~\ref{tab:procpairs}. We configured gem5 with four cores; two cores share an L1 and L2 cache, the other two cores share one additional L1 and L2, while L3 is shared by all cores. Isolated execution can use two ways per set.
We isolated each member of the two mixes (the first eight bars in Figure~\ref{fig:2ovh-4}), while the other three processes were running normally.
Each isolated process has an overhead similar to that reported in the two-process mix experiments in Figure~\ref{fig:2ovh-iso}.
Moreover, we also isolated two processes in each mix (last two columns in Figure~\ref{fig:2ovh-4}).
In this case, we measured increased overheads by up to 2 additional percentage points due to the additional competition for the \subcache.
However, those overheads are still easily tolerable given the security benefits and that they are only incurred by the isolated execution.

\subsection{Hardware and Memory Overhead} \label{sec:area}
\sys requires additional hardware and memory for the fully-associative lookup of the \subcache entries. We implemented the RTL for \sys and evaluated it for the hardware overhead for different number of isolated cache ways as shown in Table~\ref{tab:hw}, irrespective of which cache levels this is applied to.
While the overhead of the additional hardware is non-negligible, it is reasonable for a fully-associative cache lookup.
Nevertheless, it diminishes in perspective with an 8-core Xeon Nehalem~\cite{intel-xeon} of 2,300,000,000 transistors, for example. The logic overhead of \sys for 2048 fully-associative ways lookup is estimated at 1,833,200 transistors (NAND2X1 count $\times$ 4) which is 0.07\% overhead to the Xeon Nehalem.
For an 8-way 128-set cache, the memory overhead in our PoC for fully-associative mapping is 7 additional tag bits + 4 \idid bits per cache way.
With respect to access latencies, the exact timing latency of lookups will eventually depend on the circuit routing but, in principle, for a parallel content-addressable memory lookup (as in our hardware PoC), accesses are performed in 2 clock cycles.

\begin{table}[tb]
	\centering

	\begin{tabular}{>{\sffamily}c>{\ttfamily}c>{\ttfamily}c}
	\toprule
	\rmfamily\textbf{\nsecure} & \rmfamily\textbf{NAND2X1 Gates} & \rmfamily\textbf{Memory Overhead (Kb)}\\
	\midrule
	32 & 6114  & 0.34  \\
	64 & 12219  & 0.68 \\
	128 & 24563  & 1.3  \\
	256 & 48796  & 2.75  \\
	512 & 97830  & 5.5  \\
	1024 & 201792 & 11    \\ 
	2048 & 458300  & 22     \\
	\bottomrule

	\end{tabular}
	\caption{Logic and memory overhead estimates for fully-associative lookup of 46-bit addresses for different numbers of isolated cache ways (in any cache level).}
	\label{tab:hw}
\end{table}

%% file: discussion.tex
%!TEX root = main.tex

\section{Discussion} \label{sec:discuss}
\textbf{Design and Implementation Aspects.}
\sys relies on a random-replacement cache policy combined with full-associativity to provide its dynamic isolation guarantees. The implementation of the random replacement policy is delegated to the hardware designer and considered an orthogonal problem. Cryptographically-secure pseudo-random number generators (CSPRNG) or even true hardware random number generators can be used and the seed can be changed as often as required. The output of the CSPRNG cannot be predicted if it is seeded with secret randomness at the start of every process. When the seed is changed, re-keying management tasks such as cache flushing and invalidation for the re-mapping are not required, unlike in recent architectures~\cite{Qureshi18,scattercache}. This is because in \sys the randomness is only used for selection of the victim cache line, and not for locating existing cache lines in the \subcache. Furthermore, we emphasize that CSPRNG design and implementations are an orthogonal problem to our work.

The "soft" cache partitioning of \sys is a generic concept and can be applied, in principle, to any set-associative structure. In this work, we apply it to the L1, L2, and L3 (LLC) caches, but it can also be applied selectively to only some of these cache levels or to the TLB as well, or to only some cache levels in only one or more cores in a multi-core architecture that become dedicated for allocating isolated execution. 
The choice of which cache structures to apply this to and how many ways to isolate in the \subcache is delegated to the hardware designer, given that it is a more complex design decision with other metrics and trade-offs that come into play such as the size of the structure, power consumption, and logic overhead. 
The power consumption and timing overheads associated with building and routing a fully-associative cache lookup in VLSI are significant, but can be alleviated by leveraging emerging hybrid memory technologies such as DRAM-based caches~\cite{dram-cache} and STT-MRAM caches~\cite{sttram1, sttram2}.
In practice, applying \sys to the LLC or larger caches in general would be more expensive (in terms of hardware) than L1 and L2 caches, and strict partitioning might be applied instead for the LLC. 
Nevertheless, \sys can be, in principle, applied to sliced Intel LLCs. In each slice, a number of cache ways (\subcache) is reserved for isolated execution. Any mapping from the \idid to the LLC slices can be used, such that lines from a particular \idid are allocated to a specific slice. Fully-associative lookups are thus only be performed on the \subcache portion of a single slice, thus reducing the performance overheads and allowing scaling to high-core-count processors. The slice-mapping would be based only on the \idid, and thus it would not leak any information about the data address or value.

Other design decisions in \sys include the number of bits designated for \idid and thus the maximum number of concurrent isolation domains supported (see Section~\ref{sec:hw-cache}). To support more isolation domains (not concurrently) than the hardwired maximum, the cache lines of one domain can be flushed by the kernel or microcode at context switching while the next domain is switched in and is re-assigned the available \idid. Nevertheless, supporting too many isolation domains will result in increased cache utilization, and the overall performance will suffer. This is in line with conventional cache behavior, but is aggravated in \sys because isolated execution is only allowed to utilize the \subcache portion. However, this violates our working assumption \ref{itm:minority} that only the minority of the workload requires cache-level isolation.

We emphasize that cache-based side-channel leakage directly results from the design of the cache microarchitecture and, thus, it is reasonable to investigate the fundamental microarchitectural designs of caches for upcoming processor designs. While this does not address the problem for legacy systems, it provides an exploratory ground of ideas for upcoming processor designs. \sys is architecture-agnostic and can be integrated with any processor architecture (we simulated it for x86 and implemented it for RISC-V). It is also compliant with any set-associative cache architecture independent of its hierarchy and organization, and whether it is virtually or physically indexed since no indexing is involved.

\textbf{Intra-Process Isolation Support.} 
\sys can also be extended, in principle, to provide \emph{fine-grained} run-time configuration of the isolation domain \emph{within} a process, e.g., between different threads within the same process.
Besides kernel support, this requires an instruction extension to enable isolation of particular code regions or threads to different \idids or disable isolation altogether at run-time (reset its run-time \idid to all-zero).
However, this requires the developer to identify and annotate security-sensitive code regions. 
Nevertheless, this is useful in practice since a process might not require cache-based side-channel resilience for its entirety but only for sensitive code such as cryptographic computations.
This is a more generalizable approach that is easier and more directly applicable than implementing leakage-resilient variants for security/privacy-sensitive computations.

\textbf{Deployment Assumptions.} \sys assumes any TEE or trusted computing environment that is leveraged in compliance with their original design intent, i.e., that the much larger portion of the execution workload is not security-critical and only a smaller portion is security-critical and isolated in an \sd (\ref{itm:minority}).  
Otherwise, if the workload is equally balanced, the isolated execution subset would be restricted to a smaller partition of the cache and would incur a more than tolerable performance degradation especially if it is cache-sensitive. For \sys to be optimally advantageous, the workload distribution and allocation must be performed by the administrator such that the right balance of overall security and performance is achieved, as shown by the performance results in Section~\ref{sec:perfeval}.

%% file: related.tex
%!TEX root = main.tex
\section{Related Work} \label{sec:rel}
We describe next the state of the art in existing defenses and their shortcomings that \sys overcomes. 
\vspace{-.6em} % orhpan avoidance

\subsection{Partitioning}
Cache partitioning allocates to each process or security domain a separate partition of the cache, hence guaranteeing strict non-interference. Both software-based~\cite{Godfrey03, Kim12, Zhou16, Liu16} and hardware-based~\cite{Wang07,Wang16,Gruss17,Kiriansky17} partitioning schemes have been proposed in recent years, where partitioning is either process-based or region-based. 

\textbf{Process-based partitioning.} Godfrey~\cite{Godfrey03} implements process-based cache partitioning using page coloring on Xen, which incurs a prohibitive performance overhead with increasing number of processes.
SecDCP~\cite{Wang16} is a way-partitioning scheme where each application is assigned a security class and cache partitioning between the security classes is dynamically managed according to the cache demand of non-secure applications. SecDCP is not scalable; selective cache flushing and repartitioning is required if the number of security classes exceeds that of allocated partitions and it may perform worse than static partitioning. Furthermore, both schemes do not support the use of shared libraries. CacheBar~\cite{Zhou16} periodically configures the maximum number of ways allocated to each process which unfairly impacts performance and cache utilization, and does not scale well with the number of security domains. DAWG~\cite{Kiriansky17} partitions the caches where different processes are assigned to different protection domains isolating cache hits and misses. The aforementioned schemes incur the performance overhead for the entire code, whereas \sys only enables side-channel resilience and the resulting performance overhead only for the isolated execution.
 
Sanctum~\cite{costan2016sanctum} protects TEEs by flushing private caches whenever the processor switches between enclave mode and normal mode and partitioning of the LLC and assigning to each enclave a static number of sets. Sets allocated to an enclave can be used exclusively by the enclave and cannot be utilized by the OS.
On the contrary, \sys allows for a flexible and dynamic sharing of cache resources between processes (thus improving performance), while preserving cache side-channel resilience for isolated execution.

Many cache partitioning and allocation schemes~\cite{Qureshi06,Jaleel08,Liu09,Xie09,Sanchez12} have been proposed that focus on cache allocation mechanisms aiming to improve performance for multi-core caches. However, such schemes do not provide security guarantees. \sys  addresses the security/performance trade-off by providing a configurable means to enable the side-channel resilience only for isolated execution while providing non-isolated execution with unaltered performance. 

\textbf{Region-based partitioning.} These approaches split the cache into a secure partition reserved for security/privacy-critical memory pages and a non-secure partition for the remaining memory pages. STEALTHMEM~\cite{Kim12} uses page coloring where several pages are colored and reserved for security-sensitive data and they remain locked in cache. CATalyst~\cite{Liu16} leverages Intel's CAT (Cache Allocation Technology)~\cite{intel-cat} to divide the cache into secure and non-secure partitions and uses page coloring within the secure partition to isolate different processes' cache accesses to these pages. PLcache~\cite{Wang07} locks cache lines and allocates them exclusively to particular processes such that the cache line can only be evicted by its process. However, overall performance and fairness of cache utilization are strongly impacted as the protected memory size increases in relevance to the total cache capacity. Moreover, with PLcache an attacker process may still infer the victim's memory accesses by observing that it is unable to access or evict cache lines (locked by a victim process) from a particular cache set.

Cloak~\cite{Gruss17} uses hardware transactional memory, such as Intel TSX~\cite{intel-sw}, to protect sensitive computations by pre-loading the security-critical code and data into the cache at the beginning of the transaction and any cache line evictions are detected by the transaction aborting. Cloak incurs prohibitively high performance overhead for memory-intense computations and requires the developer's strong involvement to identify and instrument security-sensitive code and split it into several transactions. Recent works have also explored the LLC inclusion property for defense schemes such as RIC~\cite{Kayaalp17} and SHARP~\cite{Yan17}. However, both are architecture-specific, RIC requires coherence protocol modifications and cache flushing on thread migration, while SHARP requires modifications to the \emph{clflush} instruction. \sys, however, is architecture-agnostic, and does not require cache flushing or modifications to coherence protocols or the \emph{clflush} instruction. 

\subsection{Randomization}
Introducing randomization involves introducing noise or deliberate slowdown to the system clock to hinder the accuracy of timing measurements as in FuzzyTime~\cite{Hu91} and TimeWarp~\cite{Martin12}. These techniques can only defeat attacks which rely on measuring access latency, but cannot prevent other attacks such as alias-driven attacks~\cite{Guanciale16}. They compromise the precision of the clock for the remaining workload, thus affecting functionality requirements.

RPCache~\cite{Wang07} randomizes the mapping of all memory lines of a protected application at a per-set granularity from their actual cache set to a randomly mapped cache set, by using a permutation table. NewCache~\cite{Newcache16} randomizes the mapping at a per-line granularity using a Random Mapping Table. Both RPCache and NewCache schemes do not scale well with the number of lines in the cache (not applicable for larger LLCs) and the number of protected domains. Random Fill Cache~\cite{Liu14} mitigates only reuse-based cache collision attacks by replacing deterministic fetching with randomly filling the cache within a configurable neighborhood window whose size impacts the performance degradation incurred. It does not scale well with an increasing TEE size.

Time-Secure Cache~\cite{Trilla18} uses a set-associative cache indexed with a keyed function using the cache line address and Process ID as its input. However, a weak low-entropy indexing function is used, thus re-keying is frequently required followed by cache flushing which requires complex management and impacts performance. CEASER~\cite{Qureshi18} also uses a keyed indexing function but without the Process ID, thus also requiring frequent re-keying of its index derivation function and re-mapping to limit the time interval for an attack. A concurrent work, ScatterCache~\cite{scattercache}, uses keyed cryptographic indexing that depends on the security domain, where cache set indexing is different and pseudo-random for every domain but consistent for any given key. Thus, re-keying may still be required at time intervals to hinder the profiling and exploitation efforts of an adversary attempting to construct and use an eviction set to collide with the victim access of interest. \sys, on the other hand, leverages randomization by disabling set-associativity altogether and using random replacement for isolated execution. Every given memory address can be cached in any of the available \subcache ways and placement is random and unpredictable; it varies randomly every time the same memory line is brought in cache.

%% file: conclusion.tex
%!TEX root = main.tex

\section{Conclusion} \label{sec:conc}
In this paper, we proposed a generic mechanism for flexible and "soft" partitioning of set-associative memory structures and applied it to multi-core caches, \CHANGED{which we call} \sys. \sys effectively thwarts contention-based and access-based cache attacks by selectively applying side-channel-resilient cache behavior only for code in isolated execution domains (e.g., TEEs). Meanwhile, non-isolated execution continues to utilize unaltered and conventional cache behavior, capacity and performance. 
This addresses the persistent performance/security trade-off with caches by providing the additional side-channel resilience guarantee, and the resulting performance degradation, only for the security-critical execution subset of the workload (usually isolated in a TEE) by eliminating the fundamental causes of these attacks.
We evaluated \sys with the SPEC CPU2006 benchmark and show a performance overhead of up to 5\% for isolated execution and no overhead for the non-isolated execution.

%% file: ack.tex
%!TEX root = main.tex

\section*{Acknowledgments}
\noindent We thank our anonymous reviewers for their valuable and constructive feedback. We also acknowledge the relevant work of Tassneem Helal during her bachelor's thesis. This work was supported by the Intel Collaborative Research Institute for Collaborative Autonomous \& Resilient Systems (ICRI-CARS), the German Research Foundation (DFG) through CRC 1119 CROSSING P3, and the German Federal Ministry of Education and Research through CRISP. 